\documentclass[sigconf]{acmart}

\AtBeginDocument{%
  \providecommand\BibTeX{{%
    \normalfont B\kern-0.5em{\scshape i\kern-0.25em b}\kern-0.8em\TeX}}}

\begin{document}

\pagestyle{empty}

\title{openwifi CSI fuzzer for authorized sensing and covert channels}

\author{Xianjun Jiao, Michael Mehari, Wei Liu, Muhammad Aslam, Ingrid Moerman}
\email{{xianjun.jiao/michael.mehari/...}@ugent.be}
\affiliation{%
  \institution{IDLab, Gent University - imec}
  \city{Gent}
  \country{Belgium}
}

\renewcommand{\shortauthors}{Xianjun Jiao, et al.}

\begin{abstract}
CSI (Channel State Information) of WiFi systems contains the environment channel response between the transmitter and the receiver, so the people/objects and their movement in between can be sensed. To get CSI, the receiver performs channel estimation based on the pre-known training field of the transmitted WiFi signal. CSI related technology is useful in many cases, but it also brings concerns on privacy and security. In this paper, we open sourced a CSI fuzzer to enhance the privacy and security of WiFi CSI applications. It is built and embedded into the transmitter of openwifi, which is an open source full-stack WiFi chip design, to prevent unauthorized sensing without sacrificing the WiFi link performance. The CSI fuzzer imposes an artificial channel response to the signal before it is transmitted, so the CSI seen by the receiver will indicate the actual channel response combined with the artificial response. Only the authorized receiver, that knows the artificial response, can calculate the actual channel response and perform the CSI sensing. Another potential application of the CSI fuzzer is covert channels based on a set of pre-defined artificial response patterns. Our work resolves the pain point of implementing the anti-sensing idea based on the commercial off-the-shelf WiFi devices.
\end{abstract}

\settopmatter{printacmref=false}
\renewcommand\footnotetextcopyrightpermission[1]{} 


\maketitle

\section{Introduction}
CSI sensing \cite{ma2019wifi} based on WiFi signal has been researched for years, and more CSI applications are being developed with the aid of ML (Machine Learning). For example, the gesture recognition \cite{ahmed2020device}, home surveillance and counting the number of people \cite{yoo2020privacy} for social distancing during the COVID-19 pandemic.

CSI is the result of channel estimation based on the pre-known training field in the WiFi signal. It is the key information for signal demodulation, so it is available in all WiFi receivers. Initially, only  an SDR (Software Defined Radio) receiver could perform and expose channel estimation based on the target WiFi signal. More recently, several COTS (Commercial off-the-shelf) WiFi chips support the CSI information extraction, such as ath9k (in many devices) \cite{xie2018precise} and bcm43455c0 (in the Raspberry PI B3+/B4) \cite{gringoli2019free}. So, low cost and large scale CSI collection becomes possible, and this boosts the CSI related research further.

Meanwhile the CSI based technology brings privacy and security concerns, for example deciphering sensitive keystrokes \cite{meng2019revealing}, because anyone can capture the WiFi signal in the air. To prevent probing the environment based on the WiFi signal, a direct solution is adding artificial channel response inside the transmitter. Unfortunately, none of the COTS WiFi chips support this feature, and therefore this can only be implemented in a SDR transmitter \cite{cominelli2021ieee}.

Openwifi \cite{jiao2020openwifi} is an open source full-stack WiFi implementation based on FPGA (Field Programmable Gate Array). It operates in the same way as the COTS WiFi chip, so the usual Linux network and WiFi tools can be used together with openwifi. Currently, openwifi supports IEEE 802.11a/g/n. In this paper, we design and embed the CSI fuzzer module in openwifi to generate artificial channel response at the transmitter, thanks to the reprogrammability of FPGA. The usual WiFi fuzzer changes bits in the packet, while our CSI fuzzer affects the CSI result at the receiver by changing the signal at the transmitter. The work is started in the ORCA 3rd Opencall experiment CSI MURDER \cite{cominelli2021ieee}, it has been refactored and released as open source \cite{openwifigithub}. In this demo, we show that the artificial channel response applied to the transmitted signal has only a minor impact on the WiFi link performance.

\section{Implementation of CSI fuzzer}

\begin{figure}[h]
  \centering
  \includegraphics[width=\linewidth]{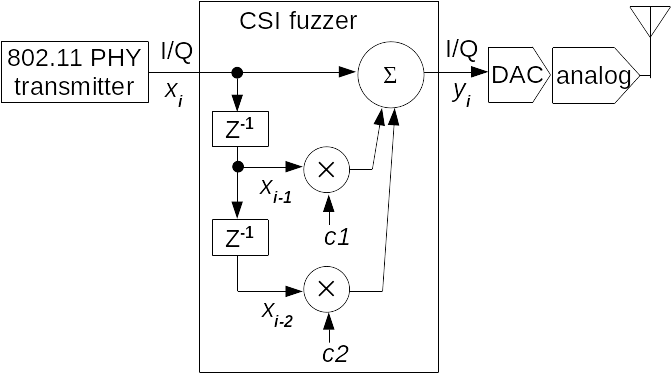}
  \caption{CSI fuzzer principle}
  \Description{CSI fuzzer principle}
  \label{fig:fuzzer-principle}
\end{figure}

Fig.~\ref{fig:fuzzer-principle} shows the design of the openwifi CSI fuzzer. It is a 3 tap FIR (Finite Impulse Response) filter between the 802.11 PHY transmitter and DAC (Digital to Analog Converter). It imposes CIR (Channel Impulse Response) \begin{math}[1, c1, c2]\end{math} to the signal. For hardware simplicity, {\it c1 c2} are complex numbers  with only real or imaginary parts and are limited in the range of \begin{math}[-0.5, 0.5)\end{math} or \begin{math}[-0.5i, 0.5i)\end{math}. They can be set via the openwifi user space tool: {\it sdrctl}. In the fuzzer, the original I/Q sample goes directly to the DAC without extra delay regardless of the delayed and attenuated versions. In this way, the original SIFS (Short Inter Frame Spacing) related real-time operation, such as the fast acknowledgment and RTS/CTS, is still guaranteed. Thanks to openwifi’s full stack, we can test the impact of the CSI fuzzer while running real WiFi traffic.

\section{Recover the CSI at the authorized receiver}

Formula (~\ref{eq:ofdm-orig}) shows the received (noise is not considered) frequency domain WiFi signal on subcarrier \begin{math}k\end{math} in the traditional OFDM (Orthogonal Frequency-Division Multiplexing) signal model.
\begin{equation}
  r(k) = \underbrace{H_{env}(k)}_{CSI(k)}{\cdot}s(k)
  \label{eq:ofdm-orig}
\end{equation}
Where \begin{math}s(k)\end{math} and \begin{math}H_{env}(k)\end{math} are transmitted signal and environment channel response on subcarrier \begin{math}k\end{math}.
The \begin{math}CSI(k)\end{math} (channel estimation result) in the receiver will be the estimation on \begin{math}H_{env}(k)\end{math}, which can be used for sensing.

According to the Digital Signal Processing theory, when a FIR filter with impulse response \begin{math}h\end{math} is applied to the signal, the equivalent frequency domain model is multiplying the frequency response \begin{math}H = DFT(h)\end{math} with the corresponding frequency domain signal, where DFT means Discrete Fourier Transform. So, the artificial CIR \begin{math}[1, c1, c2]\end{math} imported by the CSI fuzzer (FIR filter) leads to the new signal model as in the formula (~\ref{eq:ofdm-fuzzer}) for the receiver.

\begin{equation}
  r(k) = \underbrace{H_{art}(k){\cdot}H_{env}(k)}_{CSI(k)}{\cdot}s(k)
  \label{eq:ofdm-fuzzer}
\end{equation}
Where \begin{math}H_{art}(k)\end{math} is the artificial channel response added by the CSI fuzzer. Now the \begin{math}CSI(k)\end{math} will be the estimation on \begin{math}H_{art}(k){\cdot}H_{env}(k)\end{math}. To recover the estimation on \begin{math}H_{env}(k)\end{math}, the receiver just needs to do \begin{math}CSI(k)/H_{art}(k)\end{math}. For the authorized receiver, \begin{math}H_{art}(k)\end{math} can be known by \begin{math}DFT([1, c1, c2, \text{zero padding to DFT size}])\end{math}.

\section{Test result and discussions}
The hardware platform used in this demo is Zedboard + FMCOMMS2. The former contains the Xilinx Zynq 7z020 SoC chip with FPGA running openwifi design and embedded ARM processor running Linux, and the latter is the RF (Radio Frequency) board based on AD9361 that includes ADC/DAC and analog circuit till the antennas. The platform (with openwifi design running) acts as WiFi AP (Access Point), and an iPhone7 acts as client connecting to the openwifi AP. The WiFiPerf Endpoint APP (version 1.4.1) runs as the iperf TCP server on the iPhone to receive the TCP traffic from the openwifi AP. 

\begin{figure}[h]
  \centering
  \includegraphics[width=\linewidth]{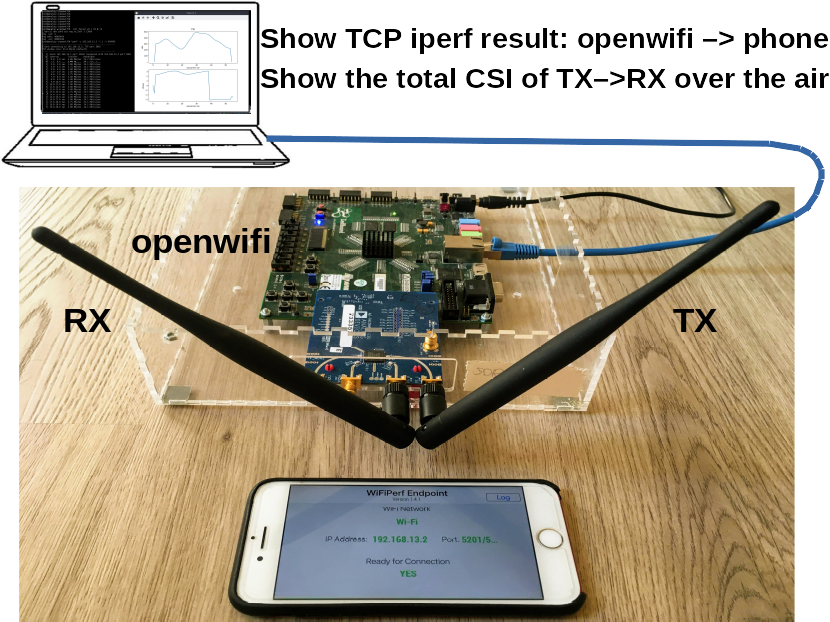}
  \caption{CSI fuzzer demo setup}
  \Description{CSI fuzzer demo setup}
  \label{fig:fuzzer-demo}
\end{figure}

Thanks to the full-duplex capability and CSI extraction feature of openwifi, we can monitor the channel response between the RX and TX antennas by receiving AP’s own packets through its RX path. Due to the half-duplex nature of the WiFi protocol, the AP’s receiver also receives incoming WiFi packets from the client when the AP is not transmitting. So, the self CSI monitoring and normal AP-client traffic can run simultaneously. Fig.~\ref{fig:fuzzer-demo} shows the setup.

\begin{figure}[h]
  \centering
  \includegraphics[width=\linewidth]{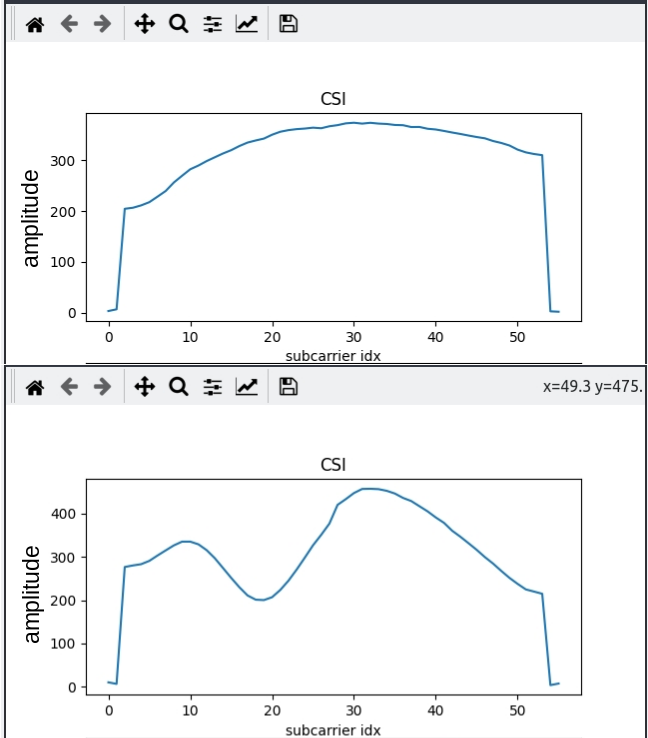}
  \caption{Monitored CSI loopback over the air. Top: CSI fuzzer off; Bottom: CSI fuzzer on}
  \Description{Monitored CSI loopback over the air. Top: CSI fuzzer off; Bottom: CSI fuzzer on}
  \label{fig:fuzzer-off-on}
\end{figure}

Two cases, with the CSI fuzzer on and off respectively, are tested while TCP iperf is running from the openwifi AP to the iPhone client. Fig.~\ref{fig:fuzzer-off-on}  shows the real-time CSI captured by openwifi receiver while TCP traffic is transmitted. In the figure’s CSI fuzzer on case, \begin{math}c1\end{math} and \begin{math}c2\end{math} are \begin{math}0.35i\end{math} and \begin{math}0.1\end{math}. If the authorized receiver knows this CIR of \begin{math}[1, 0.35i, 0.1]\end{math} added by the fuzzer, it can remove it and retrieve the environment channel response -- the original CSI.

When the CSI fuzzer is off, the CSI reflects the channel response of the very close RX-TX coupling and non-ideal analog filter effect. When the CSI fuzzer is switched on, the obvious CSI distortion (frequency selectivity) can be observed due to the additional CIR imposed by the FIR in the transmitter. The CSI change can also be observed by other CSI tools on the COTS WiFi chip.

When testing TCP iperf performance at different locations/distances with different \begin{math}c1\end{math} and \begin{math}c2\end{math}, no noticeable differences are observed on the throughput when the CSI fuzzer is on or off. Throughput of 22 \textasciitilde 24Mbps is achieved in both cases, which means that the frequency diversity brought by the WiFi FEC (Forward Error Correction) and frequency interleaver works well. Under some special multipath environment, when the artificial response happens to pre-boost the subcarriers that will experience deep fading in the multipath environment, the throughput with CSI fuzzer is even higher than the case without CSI fuzzer.

The speed of the CSI fuzzer updating script {\it csi\_fuzzer.sh} is tested via the {\it time} command in the bash. It takes 12.5ms in total to apply {\it c1 c2} to the signal. The time includes composing the register value from the {\it c1 c2} (6ms) and setting FPGA register via {\it sdrctl} (6.5ms). A script {\it csi\_fuzzer\_scan.sh} is offered to demonstrate time-variant/random artificial CIR generation to fight the artificial CIR analysis via random tests.

Using the frequency domain pattern of the artificial response to deliver a covert message is difficult, because the CSI at the receiver includes the random channel response of the environment. But when the environment is static or quasi-static, a series of the artificial response along with the time could construct a pattern which can be recognised via a series of CSI at the receiver. In this way, a covert channel can be constructed.

\section{Related work}

In \cite{schulz2018teaching}, a filter-based covert channel scheme is proposed based on reverse engineering the COTS WiFi chip. But finally, it turns out that there isn't FIR filter in the transmitter chain of the COTS chip. As an alternative solution, the distorted (based on the designed FIR response) I/Q sample of the acknowledgement packet is pre-calculated and stored into the on-chip RAM connected to the transmitter DAC to generate the signal with artificial CIR. Due to the small RAM size and non-real-time packet generation, the solution has many limitations for real application.

In \cite{meng2019revealing}, the CSI obfuscation countermeasure is proposed to thwart the inference attack. The paper proposes changing the LTF (Long Training Field) to generate the artificial CSI, because the receiver extracts CSI based on the LTF. But in theory, the artificial CSI needs to be applied to the whole packet signal, not only the LTF. Otherwise, the data portion will experience different CSI compared with the LTF portion, which will cause problem for channel equalization at the receiver. For experiment, the paper uses another COTS device at a different location to emulate the changed CSI of the device under attack, because the COTS chip lacks the capability of modifying the packet at the signal level.

Above works both have difficulties to implement the anti-sensing idea based on the COTS WiFi devices. With the openwifi CSI fuzzer, this pain point can be resolved.

\section{Conclusion and future work}

In this paper, we have demonstrated that the openwifi CSI fuzzer can add an artificial channel response to the signal inside the transmitter effectively, and it is shown that the impact on the WiFi throughput is minor. Based on the openwifi CSI fuzzer design, two use cases are proposed and discussed: authorized sensing and covert channels, which could be further explored in the future. 

This initial release of the CSI fuzzer demonstrates the basic interface and capability. In the future, the impact to the WiFi performance needs to be evaluated in more diverse circumstances. The number of taps of the CSI fuzzer and quantization/bitwidth of the tap can be extended to offer bigger degree-of-freedom/randomness for stronger protection against malicious analysis. In the current implementation, the artificial CIR becomes effective immediately after the value reaches the FPGA register, which means that the CIR could change in the middle of one packet transmission. Although this rarely happens, it should be avoided in the future to not degrade the decoding performance of that single packet.

The openwifi CSI fuzzer brings the fuzzing down to the signal level for the first time, and demonstrates the full system operating with COTS WiFi devices. We believe that this open source release will open up many new research opportunities in the area of WiFi privacy and security.

\begin{acks}
Openwifi was funded by the European Commission under the Horizon 2020 Orchestration and Reconfiguration Control Architecture – ORCA project (grant no. 732174). CSI-MURDER was funded by ORCA Open Call 3 “Experimental analysis of CSI based anti-sensing techniques - CSI-MURDER”.
\end{acks}

\bibliographystyle{ACM-Reference-Format}
\bibliography{openwifi-csi-fuzzer.bib}

\end{document}